**The mechanism of secondary structural changes in Keratinocyte Growth Factor during uptake and release from a hydroxyethyl(methacrylate) hydrogel revealed by 2D Correlation Spectroscopy**

Shohini Sen-Britain[1], Wesley Hicks[2], Robert Hard[3], Joseph A. Gardella Jr[1].

[1]Department of Chemistry, State University of New York at Buffalo

475 Natural Sciences Complex, Buffalo NY, 14202

[2]Roswell Comprehensive Cancer Center, Department of Head and Neck/Plastic and

Reconstructive Surgery, 665 Elm Street, Buffalo, NY 14203, USA

[3]State University of New York at Buffalo, Jacobs School of Medicine and Biomedical Sciences,

Department of Pathological and Anatomical Sciences, 955 Main St, Buffalo, NY 14203, USA

Email:  ssen3@buffalo.edu, gardella@buffalo.edu

**Abstract**. Incomplete release profiles of protein delivery systems can be caused by unfolding and denaturation of proteins due to protein-biomaterial interactions. This paper investigates the mechanisms causing incomplete release of the wound healing protein, Keratinocyte Growth Factor (KGF), from a hydroxyethyl(methacrylate) (HEMA) hydrogel by FTIR-ATR spectroscopy and 2D correlation spectroscopy. This work characterizes KGF secondary structure reflected in the amide I region of the FTIR-ATR spectrum, and differences between active and heat denatured KGF. These results have been used to investigate the sequence of time-dependent changes in KGF at the surface of the HEMA hydrogel during uptake and release by 2D correlation spectroscopy. KGF stays active throughout the uptake process, and the KGF loop structures interact with HEMA moieties, potentially mimicking receptor-ligand interactions. KGF denatures during release via changes in the loops and unfolding of the extended strands. Our results suggest that a high affinity interaction between the HEMA hydrogel and KGF is beneficial for efficient loading of KGF into the hydrogel, but is too strong and prevents complete KGF release due to unfolding and denaturation of KGF. This work informs future material design that will target different secondary structural elements of KGF in order to preserve its native conformation and allow for complete release.

**Introduction**

      Secondary structural changes that occur in a protein due to its interaction with a material surface in a protein-releasing delivery device can result conformational changes, denaturation, and incomplete or inefficient release. Maintaining the native conformation of a protein at the biomaterial surface can decrease issues with denaturation, trapping of the protein at the material surface, and potential issues with aggregate formation that prevent the clinical translation of published protein delivery systems. We are specifically interested in developing a drug delivery system to expedite wound healing in chronic wounds by delivering Keratinocyte Growth Factor (KGF) which is a member of the Fibroblast Growth Factor (FGF) family of proteins.

      Numerous delivery systems of FGF proteins have been developed[1-5]. In addition to the role of KGF in wound healing, FGF1 is of interest in diabetes treatment[6], and FGF2 is of interest in treating bone defects[7]. However, characterization of protein conformation or activity at the material surface intended to interact with a biological matrix is rarely addressed. Characterization studies are typically limited to bulk quantification of protein using ELISA and biological response in cellular assays. The reasons for poor biological activity of the protein delivery device are bypassed, and optimization of systems is not done. In this paper, we have used the prior FTIR characterization of the homologous FGF protein bFGF[8] to guide our characterization of the changing conformation of KGF at the surface of the HEMA hydrogel over time using 2D correlation spectroscopy. Given the close homology of proteins in the FGF family[9], our characterization method can be applied towards other delivery systems involving FGF proteins.

      Traumatic injury to the areas such as the skin, eyes, and throat can cause delays in re-epithelialization. In response to traumatic injury, the epithelium initiates an autocrine intra-epithelial repair process by expressing mitogenic and motogenic proteins such as keratinocyte growth factor (KGF)[10]. Results from our laboratories have shown that exogenous addition of

KGF to *in vitro* wound models can expedite wound closure time[11]. Biomaterial-based KGF delivery is beneficial because it allows for localized, controlled, and extended delivery of protein to the wound. HEMA-based hydrogels are adhesive porous networks that are able to swell up in the presence of water and act as tissue scaffolds[12].

KGF binds the KGF-receptor, which is a specific cell surface signal transducing receptor from the tyrosine kinase family. KGF binding to the KGF receptor requires KGF to first bind heparin[13]. KGF released by our HEMA hydrogel system would bind cell surface KGF receptors of epithelial cells in a wound, leading to cell division and migration of more cells, which would decrease the time required in the wound healing process. KGF must be released from the hydrogel and remain in its native confirmation to bind KGF receptors. In this study we are specifically interested in monitoring the conformation of KGF within first few microns of the hydrogel surface that has the potential to interact with epithelium in an *in vivo* context. FTIR-ATR spectroscopy allows for the study of conformational changes in proteins that occur at the surface of the biomaterials.

Secondary structural changes in proteins can be studied by monitoring changes in the amide I region (1600-1700 cm$^{-1}$). The amide I region consists of many overlapping bands that correspond to distinct secondary structures[14]. FTIR-ATR studies of protein conformation and activity at biomaterial interfaces have been done for over 40 years, however recent advancements in resolution enhancement techniques in the past 30 years have greatly increased the amount of information regarding protein structure that can be extracted from the FTIR spectrum[14-17].

Fourier self-deconvolution of FTIR spectra consists of transforming a region of interest back to its interferogram stage, correcting for exponential decay in the corresponding Lorentzian cosine waves that are revealed, and applying the Fourier transformation to convert the spectrum back to the frequency domain, thereby recreating the original spectrum but now with significantly narrowed bandwidths. Component bands that are now revealed by Fourier self-deconvolution can then be curve fitted by selecting prominent peaks and assuming a Lorentzian curve shape[18]. Work by Byler and Susi has shown that the fractional area of the curves created by curve fitting corresponds to the percent of the corresponding secondary structure in the protein as shown by x-ray crystallography[19]. Therefore, this analysis can be used to characterize dynamic changes that occur in protein secondary structure under different conditions. For example, the changes in secondary structure between active and heat-denatured protein can be identified. Spectral characteristics of active and denatured protein can then be monitored for in the ATR spectra of proteins at interfaces[14-17].

While Fourier self-deconvolution and curve fitting offer solutions to identifying secondary structural changes, complete resolution of the contributions of overlapping bands in the amide I region is not always possible. Furthermore, tracking the sequence of dynamic changes due to external perturbations is not always possible. 2D correlation spectroscopy has emerged recently as a technique to address these issues[20-30]. 2D correlation spectroscopy results in dynamic representations of spectra collected in response to an external perturbation such as temperature, concentration, time, etc. These representations determine the sequence of changes that occur in the overlapping bands in a region of interest. Recently, several different groups have used 2D correlation spectroscopy to characterize structural changes that occur in proteins over time, or as a function of changing concentration, during adsorption to a biomaterial[24, 31, 32]. Knowledge of regions of the protein from crystallographic data likely to interact with the biomaterial surface have guided interpretation of the 2D correlation results.

This work utilizes Fourier self-deconvolution, curve fitting, 2D correlation spectroscopy, and crystallographic information to identify the structural changes that occur in KGF over time at the HEMA hydrogel surface that lead to incomplete release. Our work is the first in the literature, to the best of our knowledge, to report a mechanism for protein denaturation at a biomaterial surface using 2D correlation spectroscopy. We have shown that incomplete release is due to denaturation of KGF at the HEMA surface, and have defined the structural elements of KGF being disrupted; the loops being most important. Our results will guide us in future second-generation hydrogel designs, which will be focused on controlling material-protein interactions in a way which avoids the disruption of structural elements required for activity and correct folding.

**Experimental**

**Materials –** 2-hydroxyethyl methacrylate, (HEMA, contains ≥ 50 ppm monomethyl ether hydroquinone as inhibitor, SKU 477028), trimethylolpropane triacrylate (TMPTMA, contains 600 ppm monomethyl ether hydroquinone as inhibitor, SKU 246808), benzoin methyl ether (BME, 96%, SKU B8703), chlorotrimethylsilane (SKU 386529), phosphate buffered saline (PBS, pH 7.4, liquid, sterile filtered and suitable for cell culture, SKU 806552), and glycerol (SKU G9012) were purchased from Sigma Aldrich. Human Recombinant Keratinocyte Growth Factor (KGF) was purchased from Prospec, and the Alexa Fluor™ 488 Microscale Protein Labeling Kit was purchased from Invitrogen (catalog #A30006).

**Hydrogel synthesis –** 0.5% HEMA hydrogels were synthesized using 5 mL HEMA, 0.5 vol % TMPTMA and 0.2% BME dissolved in glycerol. Components were mixed, degassed, and injected in between silanized glass slides separated by a 1.1 mm thick Teflon spacer, and allowed to polymerize under UV light for 30 minutes. Polymerized hydrogels were removed from the glass slides and washed three times at 70°C in triply distilled water. Glass slides were silanized in a dessicator under vacuum using chlorotrimethylsilane. Hydrogels used in uptake and release assays were cut into 1x1 cm squares and dried in an oven at 70°C for one hour.

**FTIR-ATR measurements –** Spectra were acquired using a Spectrum Two™ FTIR Spectrometer equipped with a Universal ATR accessory containing a diamond/ZnSe crystal. 64 scans were acquired for each sample at 4 cm$^{-1}$ resolution. Spectrum software was used to convert spectra from transmittance to absorbance, and to perform baseline correction. All KGF solutions used in FTIR measurements were made with unlabeled KGF. Background subtraction for 62.5 nM KGF solutions obtained in PBS was done using PBS, while subtraction for hydrated HEMA gels containing KGF/PBS was done using HEMA gels that had been hydrated in PBS. Heat-denatured KGF was prepared by boiling a 62.5 nM KGF solution in PBS at 80°C for 20 minutes.

**Deconvolution and curve fitting –** The amide I region (1600-1700 cm$^{-1}$) of the acquired spectra were Fourier self-deconvoluted using Origin Pro 2017. Parameters for deconvolution were 0.5 for gamma and 0.1 for smoothing. The peak analyzer tool was used to identify hidden peaks in the deconvoluted amide I region. The multiple peak fit tool was used to fit the deconvoluted spectra assuming a Lorentzian curve fit.

**Fluorescence monitoring of KGF Uptake and Release –** KGF was labeled with Alexa Fluor$^{TM}$ 488 using the Microscale Protein Labeling Kit from Invitrogen. Fluorescence measurements were taken using a Qubit 3.0 Fluorometer. Uptake and release into and from the HEMA hydrogel was monitored by preparing cuvettes with 1 mL of labeled KGF solution in PBS. 1x1 cm oven-dried HEMA hydrogels were placed in the solution, and the decrease in fluorescence of the labeled KGF solution was monitored for 24 hours. Uptake fluorescence was corrected for water content taken up into the hydrogel over time. Release fluorescence measurements were made by placing the protein loaded hydrogel in a cuvette containing fresh PBS in a 37°C water bath, and the increase in fluorescence was monitored for 2 hours.

**FTIR-ATR monitoring of Uptake and Release –** The FTIR-ATR spectra of KGF at the HEMA surface were collected using hydrated hydrogels that had been incubated in KGF solutions for different amount of times during the uptake and release process.

**2D Correlation Spectroscopy –** Correlation analysis of the amide I region was done by cutting out the 1600-1685 cm$^{-1}$ region of the FTIR-ATR spectra, setting a fixed baseline for all spectra, calculating the Fourier self-deconvoluted spectrum, and correcting all spectra for concentration by dividing by the area under the curve. Processed spectra were then inputted in the 2Dshige version 1.3 (Shigeaki Morita, Kwansei-Gaukuin University, 2004-2005) where the reference spectrum was the average of all collected spectra, and correlation heat maps were produced in color.

**Results and Discussion**

**Section 1. KGF Secondary Structural Elements in FTIR-ATR spectra**

KGF secondary structural elements were assigned in the amide I region (1600-1700 cm$^{-1}$) of the FTIR-ATR spectrum of unlabeled KGF in aqueous solution. Peaks were assigned by Fourier self-deconvolution and comparison to existing peak assignments of bFGF (basic fibroblast growth factor)[8]. Contributions from secondary structural elements were in agreement with crystallographic data reported in the Protein Data Bank (PDB) as revealed by Lorentzian curve fitting of the amide I region. Differences between the structures of active and heat-denatured KGF were assigned by curve fitting and Fourier self-deconvolution of the respective spectra.

**Section 1a. FTIR Spectrum of KGF**

The amide I region (1600-1700 cm$^{-1}$) of proteins and growth factors contains valuable information regarding conformation and secondary structure[14]. Previous work on the FTIR spectra of FGF-family proteins was done using bFGF, and showed agreement in contributions of secondary structural elements seen in spectra in comparison to the reported crystal structure of bFGF[8]. KGF and bFGF are 38% homologous (Supplementary Figure 1) and both are growth factors containing five extended strands with reverse turns, disordered/irregular regions, a receptor binding loop, and a sixth disordered extended strand[8]. Peak assignments of bFGF were used to guide KGF peak assignments (Figure 1) due to similarities in the FTIR spectra (Table 1), sequence similarities (Supplementary Figure 1), and structural similarities reported from crystallographic data[33]. Crystallographic data reported in the PDB was collected using a Δ23 mutant due to its stability[34]. The first 14 of the 23 residues are a loop structure. Our studies have been done with the full length construct.

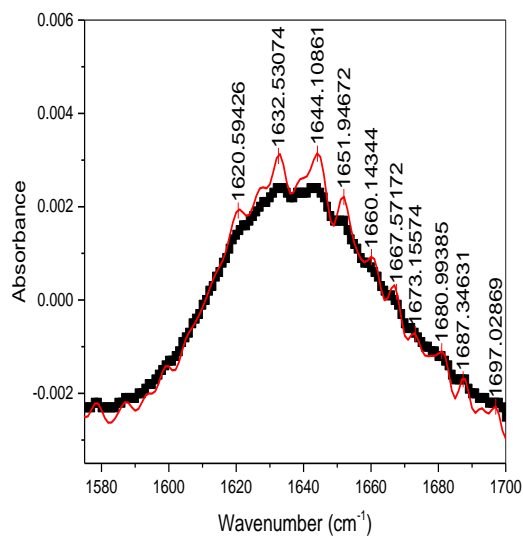

Figure 1. Original (black) and Fourier self-deconvoluted (red) spectrum of KGF in PBS

Table 1. Assignments of Secondary Structural Elements in KGF spectrum

| KGF peaks (PBS) | bFGF peaks (D$_2$O)[*] | Assignment[*] |
|---|---|---|
| 1620 | 1620 | Extended strands |
| 1632 | 1634 | Extended strands |
| 1644 | 1644 | Irregular/disordered |
| 1651 | 1650 | Loops |
| ----- | 1656 | Loops |
| 1660 | 1662 | Reverse turns |
| 1667 | 1668 | Reverse turns |
| 1672 | 1674 | Extended strands |
| 1680 | 1681 | Reverse turns |
| 1687 | 1687 | Reverse turns |
| 1696 | 1696 | Turns/Carboxyl C=O |

[*] Archives of Biochemistry and Biophysics, Vol. 285, No. 1, February 15, pp. 111-115, 1991 – The secondary structure of two recombinant human growth factors, platelet-derived growth factor, and basic fibroblast growth factor, as determined by Fourier-transform infrared spectroscopy

**Section 1b. Secondary structural contributions of KGF align with crystallographic data**

Lorentzian curve fitting of the amide I region of KGF in aqueous solution showed total contributions of secondary structural elements to be consistent with the DSSP (database of secondary structure assignments) entry for KGF in the PDB[34, 35]. Curve fitting revealed that the contribution from extended strands and reverse turns was 72.8% of the area under the amide I region, while 9.4% was irregular/disordered and 1.4% was loop structures (Figure 3, Table 2). The DSSP (database of secondary structure assignments, of the Δ23 mutant) showed that 33% of the crystal structure is extended strands, 23% is reverse turns, 38% is irregular/disordered or unassigned, and 2% is loop structures (Supplementary Figure 2). Discrepancies can be explained by the sixth disordered extended strand of KGF that appears in the region containing the extended strands and reverse turns in the FTIR spectrum resulting in an increase in the extended strand/reverse turn contribution.

The 3/10 helix that is 2% of the structure as indicated in the DSSP is a motif similar to an alpha helix and is part of the heparin-binding loop (amino acids 122-129), and previous work on bFGF indicated that alpha helix-type structures appear where loops appear in the amide I region[8]. It is important to note that the KGF construct deposited in the PDB is a Δ23 mutant, while our work has been done with the full length protein. The first 14 residues of the full length protein also form a loop. The region between 1651 and 1660 $cm^{-1}$ likely represents both this loop and the receptor binding loop, as this is seen in the FTIR spectrum of bFGF[8]. In addition to the receptor binding loop, residues 24-30 are also involved in heparin binding[34, 35]. Contributions of the extended strands and reverse turns, as well as those from the two different loops were not resolvable during curve fitting as previously published in the spectrum of bFGF. However the enhancement of the overlapping bands that allows resolution of the individual contributions is likely because the spectra were run in $D_2O$. In this study, we avoided the use of $D_2O$ in order to be able to apply the information from the characterization of KGF in aqueous solution to tracking structural changes of KGF at the surface of the hydrated HEMA hydrogel.

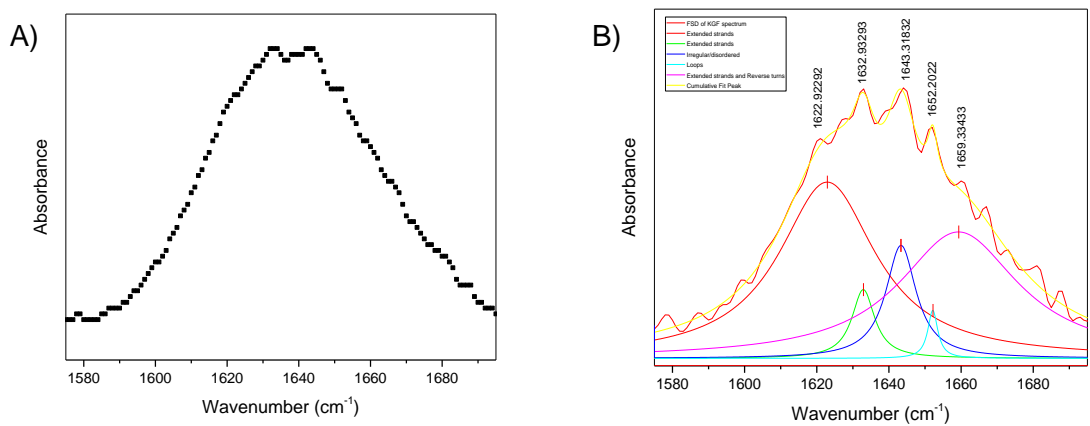

Figure 3. A) Original FTIR-ATR spectrum of KGF in PBS, B) Lorentzian curve fitting of the Fourier self-deconvoluted spectrum of KGF indicating contributions of secondary structural elements

Table 2. Percentages of secondary structural contributions in the FTIR spectrum of KGF

| Peaks used for fitting | % of Total Area in KGF FTIR Spectrum | Assignment |
|---|---|---|
| 1622 | 37.5 | Extended strands |
| 1634 | 4.4 | Extended strands |
| 1643 | 9.4 | Irregular Disordered |
| 1651 | 1.39 | Loops |
| 1660 | 30.87 | Extended strands and reverse turns |

**Section 1c. Secondary Structural Differences between Active and Heat-Denatured KGF**

In order to determine whether KGF is denaturing at the surface of the HEMA hydrogel, we modeled structural changes that occur in KGF by comparing active KGF and heat-denatured KGF. Curve fitting revealed that changes in the contributions of the extended strands (1634 cm$^{-1}$) and the irregular/disordered regions (1643 cm$^{-1}$) occur upon heat denaturation (Figure 4, Table 3). Additionally, an overall change in peak shape resulting from a change in the relative intensities of the 1634 and 1644 cm$^{-1}$ occurs. We have identified this change in peak shape as indicative of denaturation.

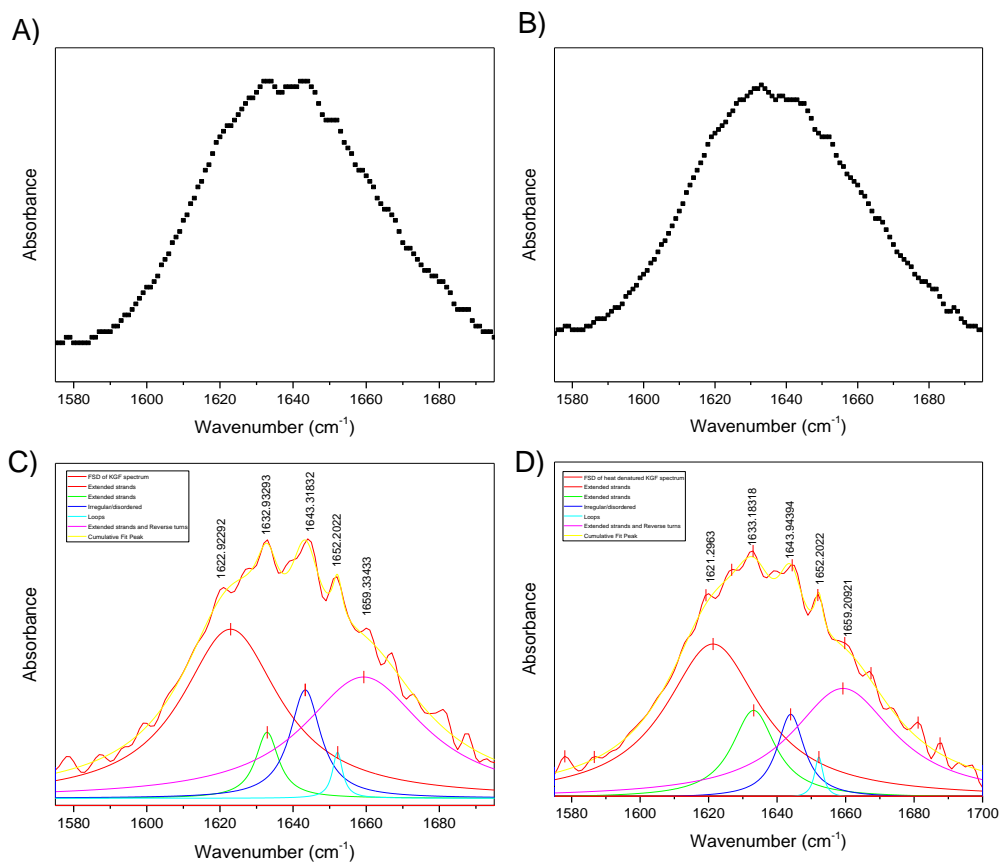

Figure 4. A) Original FTIR-ATR spectrum of active KGF in PBS, B) Original FTIR-ATR spectrum of heat denatured KGF in PBS, C) Fourier self-deconvoluted and curve fitted spectrum of active KGF, D) Fourier self-deconvoluted and curve fitted spectrum of heat denatured KGF in PBS

Table 3. Comparison of secondary structural contributions in active and heat-denatured KGF

| Peaks used for fitting | % of Total Area in Active KGF | % of Total Area in Denatured KGF | Assignment |
|---|---|---|---|
| **1622** | 37.5 | 39.5 | Extended strands |
| **1634** | 4.4 | 12.2 | Extended strands |
| **1643** | 9.4 | 7.8 | Irregular/disordered |
| **1651** | 1.4 | 1.3 | Loops |
| **1660** | 30.8 | 29.6 | Reverse turn |

**Section 2. Fluorescence monitoring of AlexaFluor 488-labeled KGF uptake and release from hydrogel**

Fluorescence monitoring of AlexaFluor 488-labeled KGF (AF488-KGF) going into and being released from the HEMA hydrogel was used to create uptake and release profiles (Figure 5). On average, ~20% of 5, 15, and 30 nM solutions of AF488-KGF were taken up into the hydrogel. The percent of initial solution concentration taken up into the hydrogel was not statistically different as indicated by one-way ANOVA (at the 0.05 level, $F_{2,2}= 0.14455$, $p > 0.05$) between the three different uptake samples. However, the raw amount of AF488-KGF taken up into the hydrogels was statistically different in the 30 nM sample when compared to the 5 and 15 nM samples (Figure 5). This result indicates the possibility of a mechanism where the formation of monolayers of KGF adsorbed onto the surface of the hydrogel or within the porous network lowers the energy of subsequent layer formation in a concentration dependent manner. These results also indicate no fixed "maximum" of KGF loading in the samples tested. Uptake into the hydrogel from 60 and 80 nM samples was also tested and resulted in ~20% uptake with a concentration dependent, statistically significant increase in the raw amount of AF488-KGF taken up into the samples. However, uptake and release results from the 60 and 80 nM samples have been omitted here due to high variability in release profiles.

Release profiles at 37°C indicate an average release of ~56% of loaded KGF calculated during generation of the uptake profile. The raw amount of AF488-KGF released was statistically higher in the 30 nM, while the percent released was statistically insignificant as indicated by one-way ANOVA between the 5 nM, 15 nM, and 30 nM samples (at the 0.05 level, $F_{2,2}= 5.05594$, $p > 0.05$). These results suggest that KGF is being incompletely released due to irreversible binding to the hydrogel network through its interactions with the HEMA moieties. This irreversible binding of KGF is likely leading to changes in secondary structure that may lead to denaturation of KGF at the HEMA surface. In an effort to design a drug delivery system that both (1) maximizes the percent of KGF released and (2) minimizes trapping and denaturation of KGF at the hydrogel surface, we have aimed to determine the interactions between KGF and the HEMA network that are leading to incomplete release profiles.  This information can inform us on future material design to avoid disrupting the implicated secondary structures of KGF causing denaturation.  Furthermore, we hypothesize that preventing KGF denaturation will increase the percent of KGF released from the hydrogel.  Uptake profiles are collected for the first two hours because previous experiments (not shown here) indicate that all KGF capable of being released from the hydrogel is released in this time frame.  However, time points beyond two hours are of interest in FTIR-ATR studies in order to understand KGF-HEMA interactions after extended periods of time.

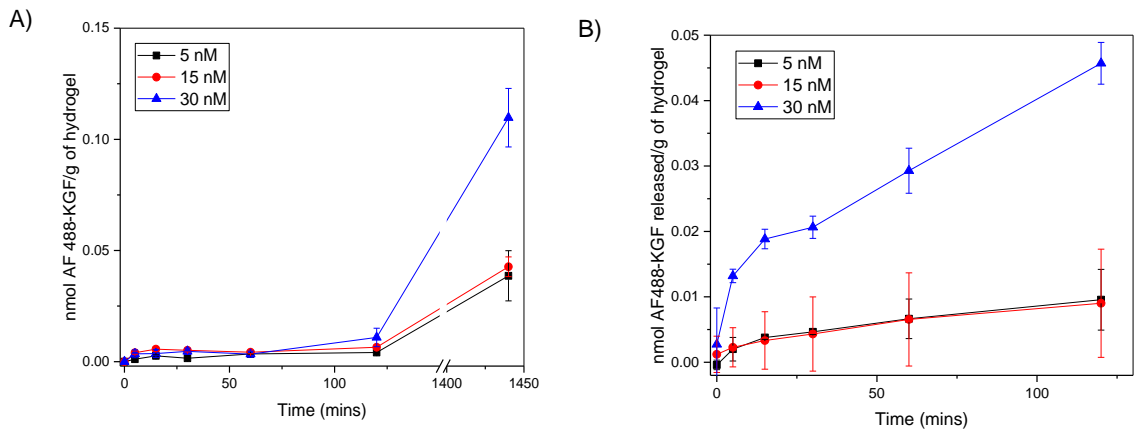

Figure 5. A) Uptake profile of 1x1 cm HEMA hydrogels in 5, 15, and 30 nM AF488-KGF solutions, B) Release profiles of 1x1 cm HEMA hydrogels loaded in 5, 15, and 30 nM AF488-KGF solutions

**Section 3.  KGF Structural Changes Revealed by 2D Correlation Spectroscopy**

**3a. Overview of 2D Correlation Analysis**

Fluorescence data of KGF uptake and release in the HEMA hydrogels indicates that KGF is likely getting trapped in or at the surface of the hydrogel network during the release process.  This trapping of KGF results in an incomplete release profile, and is potentially causing the denaturation of KGF due to interactions with the HEMA hydrogel network.  2D correlation analysis of FTIR-ATR spectra collected over time during the uptake and release process has been used to determine the changes that are occurring in the secondary structure of unlabeled KGF at the surface of the HEMA hydrogel and whether KGF is active or denatured (1) as it is being taken up into the hydrogel and (2) as it is being released from the hydrogel.  2D correlation analysis has also revealed the sequence of changes in secondary structural motifs that lead to eventual denaturation of KGF during the release process which we predict is causing incomplete release of KGF from the hydrogel.

2D correlation analysis enhances the information extracted from spectral data that has been collected as a function of an external perturbation such as time, concentration, temperature etc[36].  It is ideal for systems that contain many overlapping peaks, such as the amide I region of protein FTIR spectra.  The analysis enhances resolution of the overlapping peaks by spreading peaks in a second dimension.  Changes in selectively coupled peaks and the sequence of events over the course of the applied external perturbation can be extracted.  The signs of cross peaks in the synchronous and asynchronous 2D correlation spectra indicate the sequence of events, and cross peaks of interest are chosen by examining maxima in the reference (averaged spectrum over the course of the external perturbation) spectrum[31].

The synchronous 2D correlation spectrum depicts the correlation of simultaneous variation in spectral intensities that have been measured at two different wavenumbers $\nu_1$ and $\nu_2$.  Peaks on the diagonal correspond to the autocorrelation function of variations in spectral intensity, and off-diagonal cross peaks represent the simultaneous variation in the two wavenumbers.  A positive value on the correlation spectrum indicates that the spectral intensities at the two wavenumbers increase or decrease in the same direction, while a negative value indicates that one is increasing while the other is decreasing[31].

The asynchronous 2D correlation spectrum depicts the correlation of sequential variation in spectral intensities that have been measured at two different wavenumbers $\nu_1$ and $\nu_2$, and represents the dissimilarity of variations in spectral intensity.  When a large value is observed, it indicates that the spectral intensities at the two wavenumbers vary independently.  Independent variation in spectral intensities indicate that the signals originate from distinct moieties that respond differently to the external perturbation that has been applied[31].

The sign of cross peaks derived from the synchronous and asynchronous 2D correlation spectra can determine the sequential order of variations that occur in the spectral intensities. When the values at $(\nu_1,\nu_2)$ (1) have the same sign it indicates that changes in $\nu_1$ occurred before $\nu_2$, (2) are different it indicates that changes in $\nu_2$ occurred before $\nu_1$, and (3) if there is no cross peak in the asynchronous spectrum it indicates that changes in $\nu_1$ and $\nu_2$ occurred simultaneously[31].

### 3b. KGF Structural Regions of Interest

It is important to note that KGF has several solvent exposed regions that are likely to be involved in interactions with the HEMA hydrogel.  In Figure 6A, the extended strands (orange), reverse turns (red), disordered regions (blue), and receptor binding loop (pink) have been labeled.  The reverse turns, loop, and disordered regions are solvent exposed, while the extended strands comprise of a tightly knit hydrophobic core[33].  Figure 6B shows the structure of heparin which KGF binds with its receptor binding loop.  Heparin is comprised of carboxylic acid, sulfonic acid, hydroxyl, and sulfonamide groups.  Figure 6C shows the structure of HEMA. The presence of hydroxyl groups in HEMA make hydrogen bonding with KGF likely to occur in a way that is similar to the heparin-KGF interaction.

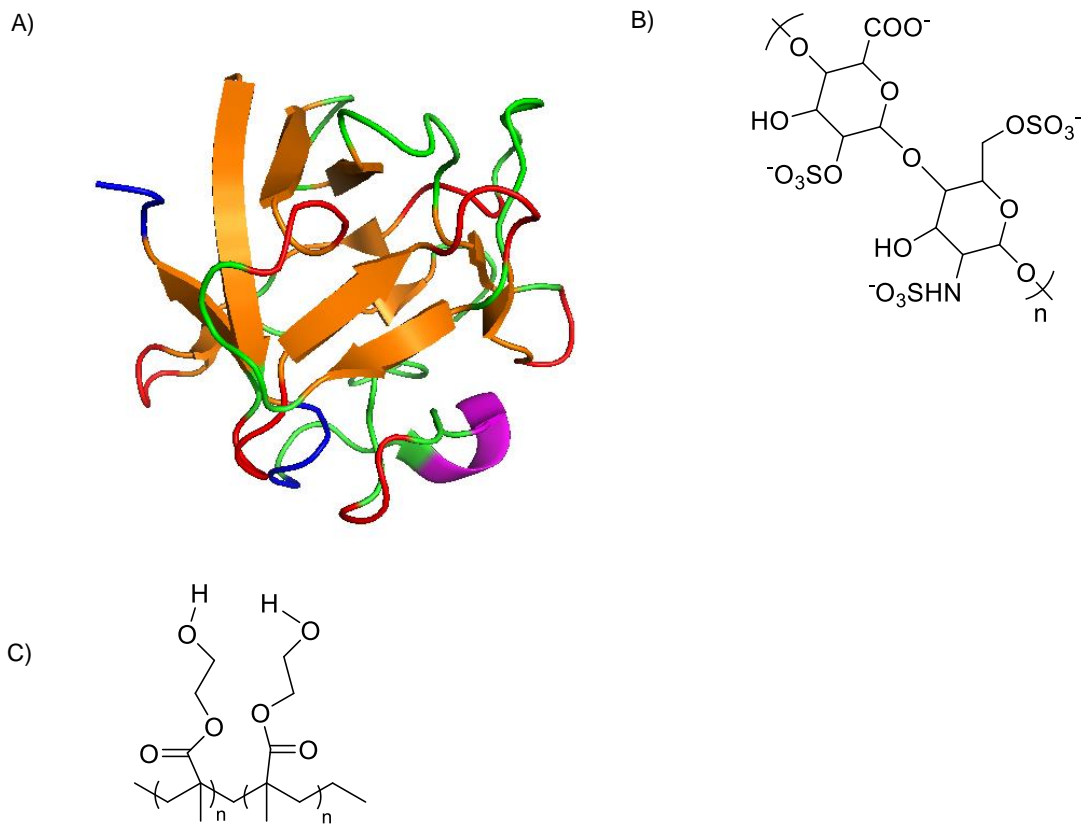

Figure 6. A) Crystal structure of KGF with secondary structural elements labeled created in PyMol: red – reverse turns, orange – extended strands, blue – disordered, pink – heparin binding loop, green – unassigned, PDB ID: 1QQK, B) Structure of heparin which is a glycosaminoglycan that binds the receptor binding loop, C) Structure of HEMA in the hydrogel

**3c. 2D Correlation Spectroscopy of KGF structural changes over time during uptake into the hydrogel**

Values of cross peaks from the synchronous and asynchronous correlation maps in Supplementary Table 1 were used to determine the sequence of structural changes that occurred in KGF upon uptake. Visual inspection of the raw spectra of KGF at the HEMA surface (Supplementary Figure 3) and comparison to the active and heat denatured spectra of KGF obtained in Section 1 indicates that KGF remains active throughout the uptake process due to the ~1:1 peak intensities at 1634 and 1644 cm$^{-1}$. Therefore, the results of this 2D correlation analysis reveal a sequence of changes in KGF secondary structure that allow KGF to remain active. A systematic increase in protein surface concentration depicted by increases in the area of the amide I region is typically seen in experiments involving measurements of protein adsorption on material surfaces[15, 16, 32]. However, this is likely not observed here due to the occurrence of both adsorption at the HEMA surface and diffusion into the porous network which would not cause a homogenous increase in protein monolayers at the surface[16]. The reference spectrum for uptake over time shows five weak maxima at 1634, 1645, 1653, 1660, and 1668 cm$^{-1}$ (reference spectrum is labeled on top of correlation map). The lack of white regions on the synchronous correlation map indicate that the changes in these moieties are highly correlated (Figure 7). The cross peaks reveal the sequence to be:

1653 (loops) ≈ 1668 (reverse turns) ≈ 1660 (reverse turns) → 1645 (irregular/disordered regions) → 1634 (extended strands)

Interpretation of the cross peaks is further simplified by examining the slice spectrum at 1653 cm$^{-1}$ (Figure 8). The slice spectrum clarifies the separation of overlapping bands and the order of intensity changes between the different bands[24]. Values of spectral intensities at 1653, 1660, and 1668 cm$^{-1}$ above zero and values of spectral intensities of 1645 and 1634 cm$^{-1}$ below zero confirm that the positive changes occurred before the negative changes. The asynchronous spectrum has been reported to be more sensitive to changes than the synchronous spectra, additional peaks are seen in the asynchronous slice spectrum (Figure 7)[24].

Of particular interest is that the interaction between KGF and HEMA starts with the loops. Signals from between 1651 and 1656 cm$^{-1}$ correspond to the receptor binding loop and the loop at residues 1-14 which are both solvent exposed. The receptor binding loop binds heparin, while residues 24-30 near the loop formed by the first 23 residues also bind heparin. Given that the loops are solvent exposed and are either responsible for binding heparin to facilitate KGF receptor binding or near a heparin-binding region, it is possible that a high affinity interaction occurs between HEMA moieties and the loops that is mimicking the interaction between the receptor binding loop/residues 24-30 and heparin. The hydroxyl bonds of HEMA may mimic the carboxylic acid, sulfonic acid, and sulfonamide moieties in heparin. The interactions with the reverse turns and irregular/disordered regions possibly occur next due to their solvent exposed character as well, and the tight hydrophobic core of the extended strands are last to interact.

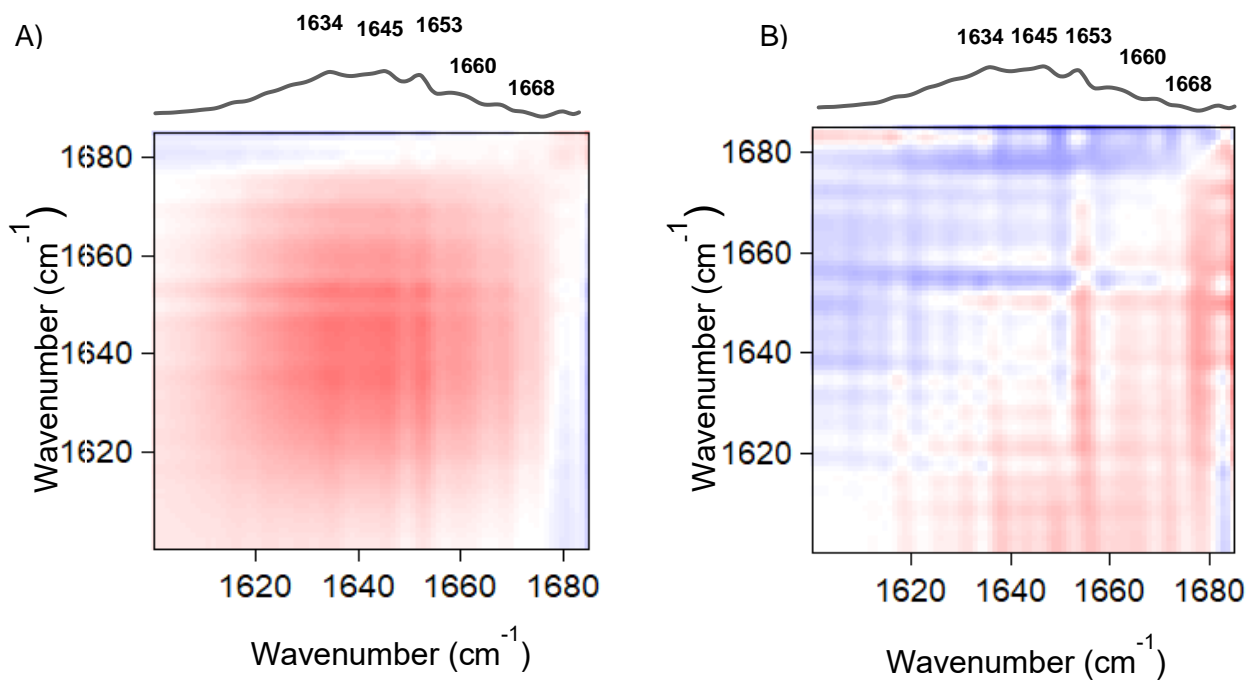

Figure 7. A) Synchronous 2D correlation map of KGF uptake into HEMA hydrogel over time with reference spectrum at top, B) Asynchronous 2D correlation map of KGF uptake into HEMA hydrogel over time with reference spectrum at top

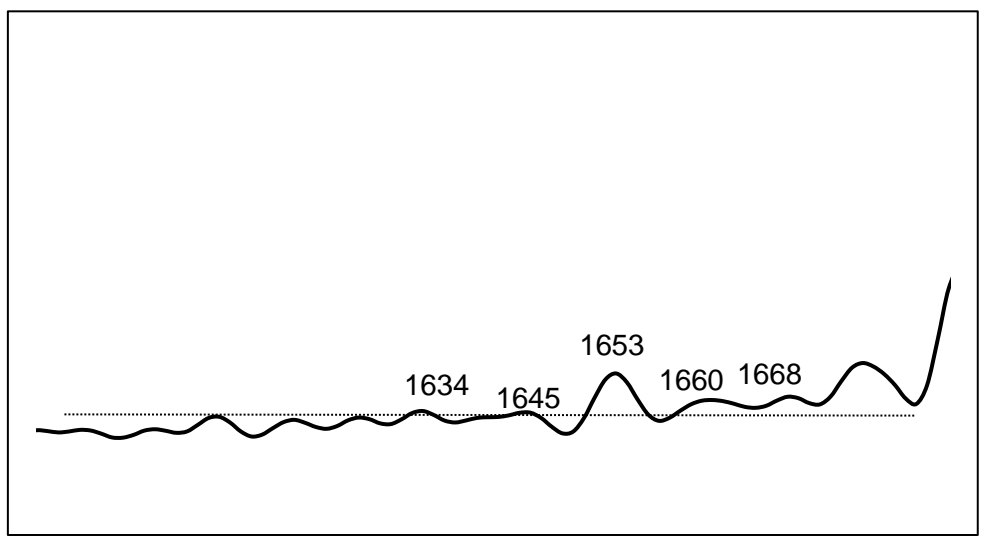

Figure 8. Asynchronous slice spectrum of KGF uptake into HEMA hydrogel over time at 1653 cm$^{-1}$ clarifies sequence of changes in amide I region

**3d. 2D Correlation Analysis of KGF at the surface of HEMA during the release process**

Visual inspection of the original KGF spectra indicate that KGF appears to be denatured by 24 hours of release at 37°C due to the large difference between the peak intensities of 1634 and 1645 cm$^{-1}$ (Supplementary Figure 4). Temperature is not expected to be the cause of denaturation here, as thermal stability of KGF up to 57°C has been reported[37]. Denaturation that is observed is therefore most likely due to the KGF-HEMA interaction at extended time points. Prior to 24 hours, the peak intensities remain ~1:1 and indicate that KGF is likely active. The values of the cross peaks from the synchronous and asynchronous correlation maps determine a sequence of changes that are therefore a mechanism leading to KGF denaturation (Supplementary Table 2). Furthermore, this result shows that KGF remains (1) trapped at the HEMA surface throughout the release process and (2) denatures due to interactions with HEMA after 24 hours. The reference spectrum generated from the spectra over time show maxima that are significantly more well-defined than the maxima seen during uptake. These maxima are at 1634, 1651, 1655, 1671, and 1680 cm$^{-1}$ (Figure 9). The cross peaks reveal the sequence to be:

1655 (loops) ≈ 1671 (extended strands) ≈ 1680 (reverse turns) → 1634 (extended strands) → 1651 (loops)

The synchronous slice spectrum at 1680 cm$^{-1}$ shows positive values of similar intensity at 1655, 1671, and 1680 cm$^{-1}$ and negative values at 1634 and 1651 cm$^{-1}$, confirming the mechanism determined by analysis of cross peaks (Figure 10). Similar to the mechanism occurring during uptake, the changes in KGF secondary structure during release also starts with involvement of the loops. The interactions between HEMA moieties and the loop are therefore both implicated in the adsorption and diffusion of KGF onto the hydrogel and into its porous network, but are also a key interaction leading to denaturation. However, the extended strands appear to be more involved in the process of KGF becoming denatured. Therefore, a high affinity interaction between HEMA and KGF is beneficial during uptake because it allows for a high loading of KGF into the hydrogel, but starts a cascade of events occurring while KGF is in or at the surface of the hydrogel that lead to unfolding and denaturation. This suggests that the extended strands may be unfolding over time in a more extensive interaction with HEMA than during uptake, which leads to trapping of KGF at the HEMA surface over time. It is likely that extensive hydrogen bonding between the hydroxyl bonds of HEMA and the extended strands lead to this unfolding.

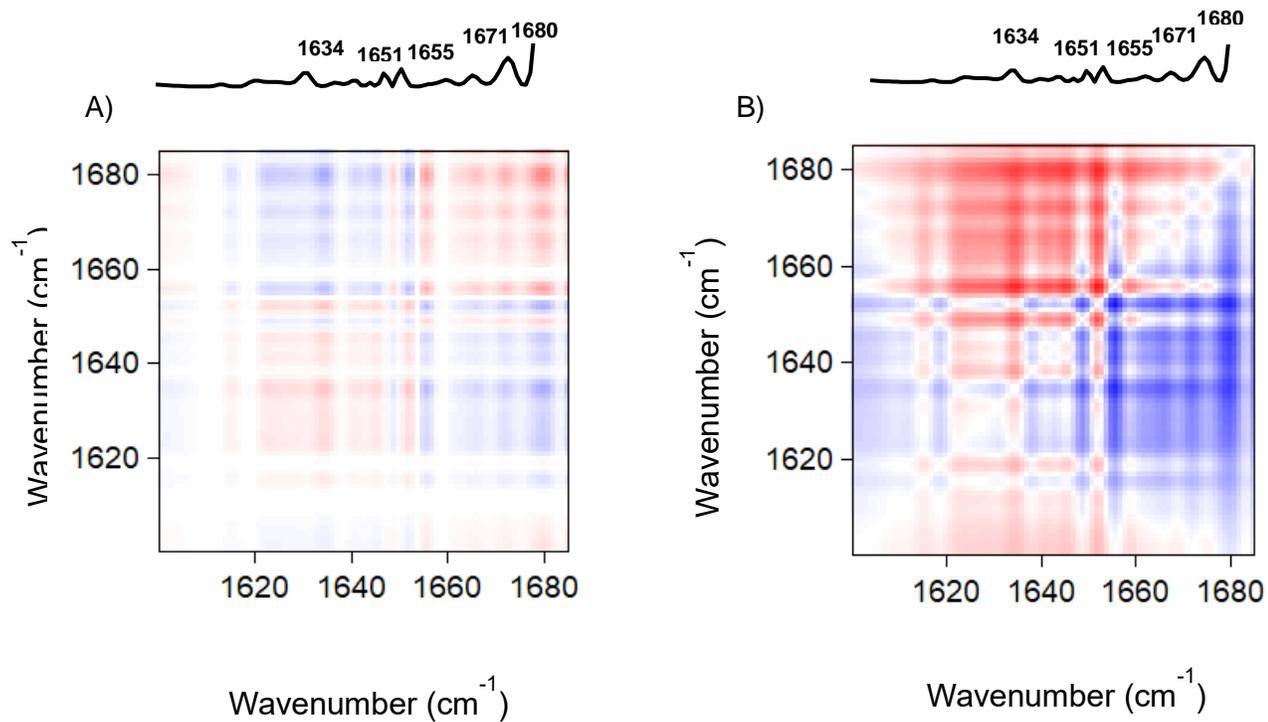

Figure 9. A) Synchronous 2D correlation map of KGF release from HEMA hydrogel over time, B) Asynchronous 2D correlation map of KGF release from HEMA hydrogel over time

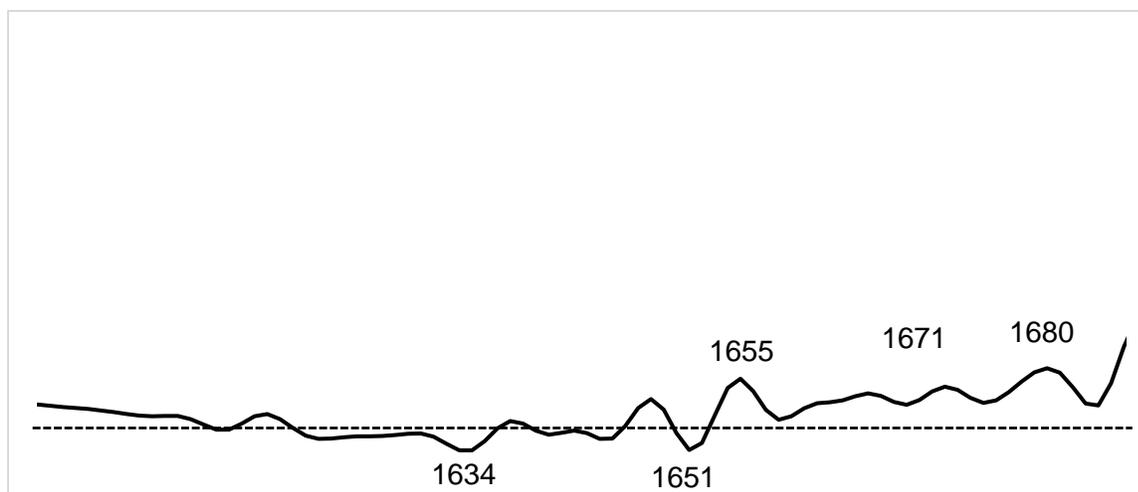

Figure 10. Synchronous slice spectrum of KGF uptake into HEMA hydrogel over time at 1680 cm$^{-1}$ clarifies sequence of changes in amide I region

**Conclusions**

To summarize our results, 2D correlation analysis has provided us with the sequence of secondary structural changes that occur over time during the uptake and release process. Prior characterization of the bFGF FTIR spectrum guided our assignment of peaks to corresponding secondary structures in the KGF FTIR spectrum. The identification of differences between the spectra of active and heat denatured KGF allowed us to determine that KGF remains active as it is taken up into the hydrogel but denatures over time during release. The interactions between HEMA moieties and the loops appear to be a driving force for (1) uptake of KGF into the hydrogel and (2) unfolding of the extended strands and reverse turns during release that lead to denaturation. Solvent exposed structures in KGF interact with HEMA prior to the buried extended strands during uptake while the extended strands play a larger role in release. This distinguishes the uptake and release mechanisms, and is likely the reason that KGF retains its activity during the uptake process and does not unfold, but denatures during release. Fluorescence tracking of Alexa Fluor™-488 labeled KGF shows that only ~60% of loaded KGF is released, and the FTIR-ATR results show that denaturation of KGF at the hydrogel surface during release is likely the cause of incomplete release.

Given the involvement of the loops as a potential driving force for KGF denaturation, the evaluation of second generation hydrogels will focus on those that interact with secondary structures of KGF not involved in receptor binding.  This is the subject of our follow-up paper, where we have investigated the role of the surface chemistry of second generation hydrogels on KGF conformation and orientation by FTIR-ATR and time-of-flight secondary ion mass spectrometry (ToF-SIMS)[38].  While the KGF-HEMA interaction is beneficial in the uptake process, it is likely that it is a high affinity interaction that hinders the release process because the biomaterial-protein interaction appears to act as an inhibitor of KGF activity.  Confirmation of this hypothesis is the subject of future *in vitro* heparin binding assays that will test the ability of HEMA-loaded KGF to bind heparin.

The characterization and study of a complex growth factor at a biomaterial interface by comparison to the existing FTIR characterization of a homologous growth factor opens up opportunities in the characterization of other systems targeting the delivery of the FGF family of proteins. The characterization done in these systems is usually limited to bulk quantification of protein using ELISA, bulk biological response in *in vivo* systems, or is nonexistent completely[1-5]. Our work here has outlined a set of FTIR experiments that can be easily done to (1) determine the folding of the protein of interest in solution when active or heat denatured, (2) determine whether these spectral characteristics are seen in the protein at the surface of the delivery system over time, and (3) use 2D correlation analysis to determine the sequence of events occurring between the protein and the biomaterial leading to the conformational changes.  The methods of analysis can then be further used to screen different materials to identify the characteristics disrupting the identified secondary structural elements of the protein, and chose a material less likely to disrupt the native folding of the protein.  Minimizing these issues may lead to the clinical translation of a higher percentage of the developed published protein delivery systems.

## Supporting Information

The supporting information is available free of charge via the internet at http://pubs.acs.org. It includes the KGF and bFGF sequence alignment, the reprint of the DSSP (database for secondary structural assignments) for KGF in deposited in the PDB, and tables containing values from asynchronous and synchronous correlation spectra, as well as original FTIR-ATR spectra of all uptake and release experiments.

## Acknowledgements

These studies were supported by funding from the John and Frances Larkin Endowment awarded to J. A. G.

**Table of Contents Image**

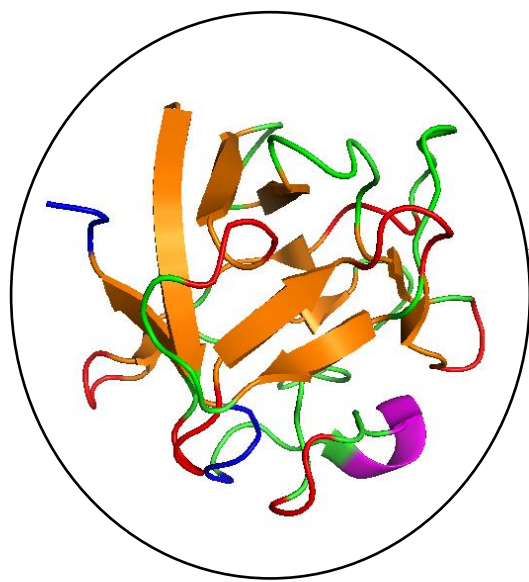
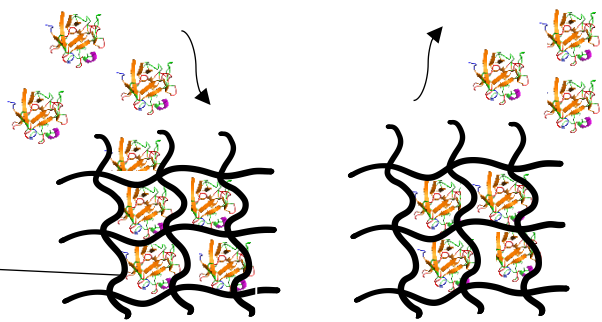
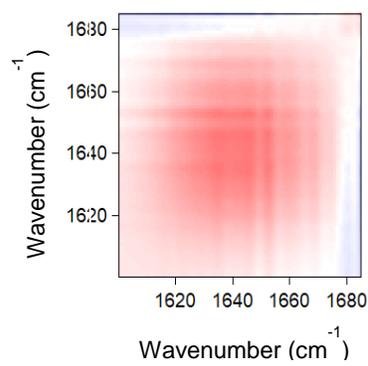
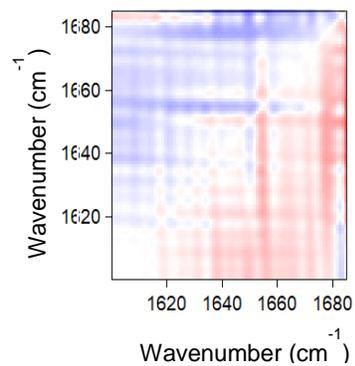

**The mechanism of secondary structural changes in Keratinocyte Growth Factor during uptake and release from a hydroxyethyl(methacrylate) hydrogel revealed by 2D Correlation Spectroscopy**

Shohini Sen-Britain[1], Wesley Hicks[2], Robert Hard[3], Joseph A. Gardella Jr[1].

[1]Department of Chemistry, State University of New York at Buffalo

475 Natural Sciences Complex, Buffalo NY, 14202

[2]Roswell Comprehensive Cancer Center, Department of Head and Neck/Plastic and Reconstructive Surgery, 665 Elm Street, Buffalo, NY 14203, USA

[3]State University of New York at Buffalo, Jacobs School of Medicine and Biomedical Sciences, Department of Pathological and Anatomical Sciences, 955 Main St, Buffalo, NY 14203, USA

Email: ssen3@buffalo.edu, gardella@buffalo.edu

**SUPPLEMENTARY INFORMATION**

```
KGF    1 SYD---------------YMEGGDIRVRRLFCRT-QWYLRIDKRGKVKGTQEMRNSYNIMEIRTVAV
bFGF   1 PALPEDGGSGAFPPGHFKDPKRLYCKNGGFFLRIHPDGRVDGVREKSDPHIKLQLQAEER

KGF   52 GIVAIKGVESEYYLAMNKEGKLYAKKECNEDCNFKELILENHYNTYASAKWTHSGGEMFV
bFGF  61 GVVSIKGVSANRYLAMKEDGRLLASKSVTDECFFFERLESNNYNTYRSRKYTS-------WYV

KGF  112 ALNQKGLPVKGKKTKKEQKTAHFLPMAI--T
bFGF 117 ALKRTGQYKLGSKTGPGQKAILFLPMSAKS
```

Supplementary Figure 1. Sequence alignment of KGF and bFGF showing 38% homology (Δ23 construct)

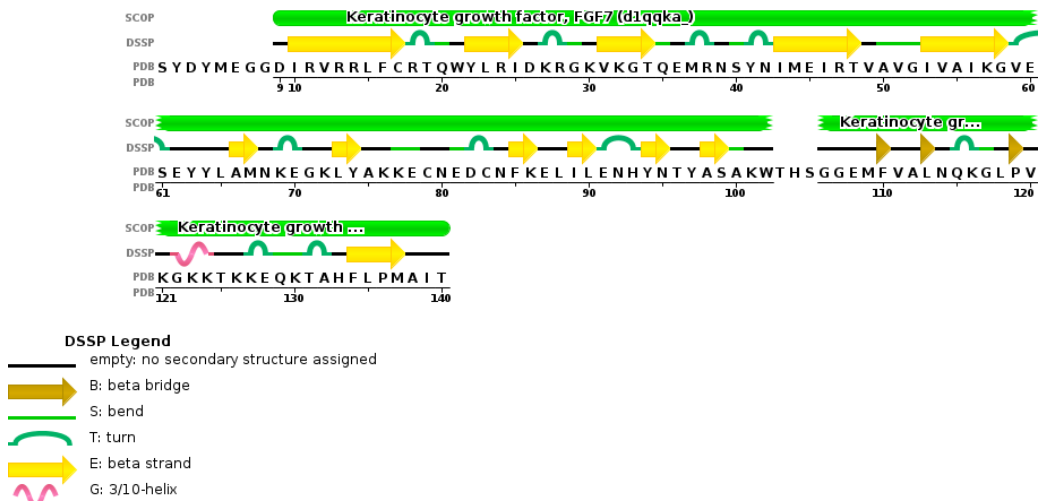

Supplementary Figure 2. Sequence chain view with DSSP information for KGF (PDB ID: 1QQK), reprinted from https://www.rcsb.org/structure/1QQK[1]
2

Supplementary Table 1. 2D-COS analysis of synchronous and asynchronous values of cross peaks for changes in KGF structure at the surface of the HEMA hydrogel during uptake

| Cross peak ($\nu_1$, $\nu_2$) | Synchronous value | Asynchronous value | Sequence |
|---|---|---|---|
| (1634, 1645) | + | - | 1634 < 1645 |
| (1634. 1653) | + | - | 1634 < 1653 |
| (1634. 1660) | + | - | 1634 < 1660 |
| (1634, 1668) | + | - | 1634 < 1668 |
| (1645, 1653) | + | - | 1645 < 1653 |
| (1645, 1660) | + | - | 1645 < 1660 |
| (1645, 1668) | + | - | 1645 < 1668 |
| (1653, 1660) | + | + | 1653 < 1660 |
| (1653, 1668) | + | + | 1653 < 1668 |
| (1660, 1668) | + | - | 1660 < 1668 |

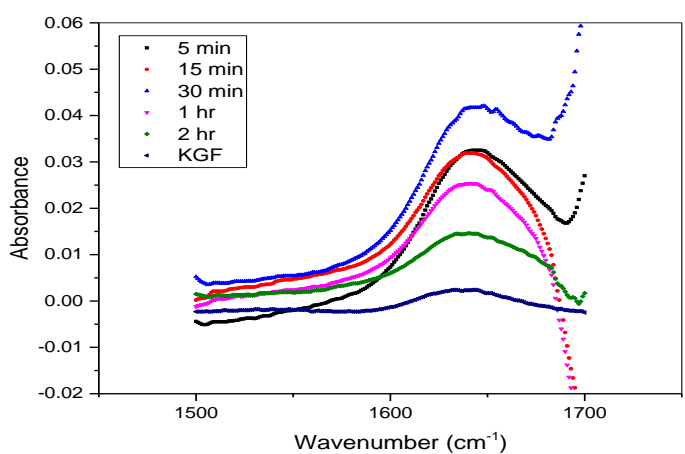

Supplementary Figure 3. Original spectra of KGF at HEMA surface during uptake and KGF in aqueous solution for comparison

Supplementary Table 2. 2D-COS analysis of synchronous and asynchronous values of cross peaks for changes in KGF structure at the surface of the HEMA hydrogel during release

| Cross peak ($v_1$, $v_2$) | Synchronous value | Asynchronous value | Sequence |
| --- | --- | --- | --- |
| (1634, 1651) | + | + | 1634 > 1651 |
| (1634. 1655) | - | + | 1634 < 1655 |
| (1634. 1671) | - | + | 1634 < 1671 |
| (1634, 1680) | - | + | 1634 < 1680 |
| (1651, 1655) | - | + | 1651 < 1655 |
| (1651, 1671) | - | + | 1651 < 1671 |
| (1651, 1680) | - | + | 1645 < 1680 |
| (1655, 1671) | + | + | 1655 < 1671 |
| (1655, 1680) | + | + | 1655 < 1680 |
| (1671, 1680) | + | + | 1671 < 1680 |

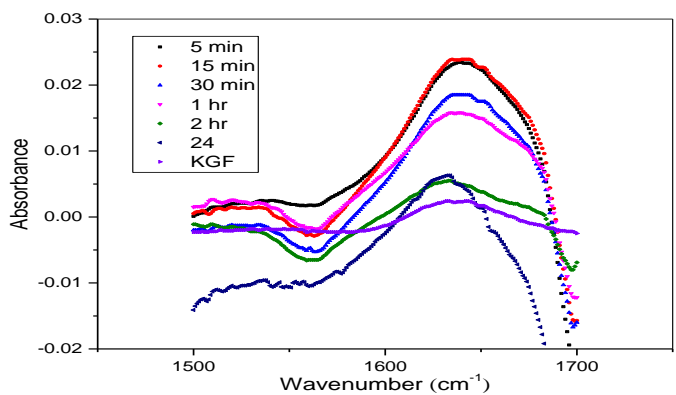

Supplementary Figure 4. Original spectra of KGF at HEMA surface during release and KGF in aqueous solution for comparison

1. Ye, S.; Luo, Y.; Lu, W.; Jones, R. B.; Linhardt, R. J.; Capila, I.; Toida, T.; Kan, M.; Pelletier, H.; McKeehan, W. L., Structural basis for interaction of FGF-1, FGF-2, and FGF-7 with different heparan sulfate motifs. *Biochemistry* **2001,** *40*, 14429-39.